\documentclass{osa-article}

\journal{osac}

\articletype{Research Article}

\usepackage[per-mode=symbol]{siunitx}
\usepackage{multirow,makecell,bm,booktabs}
\usepackage{adjustbox}

\begin{document}

\title{Towards low loss non-volatile phase change materials  in mid index waveguides}

\author{Joaquin Faneca,\authormark{1,2}  Ioannis Zeimpekis, \authormark{1} S. T. Ilie, \authormark{1} Thal\'ia Dom\'inguez Bucio,\authormark{1} Katarzyna Grabska,\authormark{1} Daniel W. Hewak,\authormark{1} and Frederic Y. Gardes\authormark{1,*} }

\address{\authormark{1}Optoelectonics Research Center, University of Southampton,  Southampton, SO17 1BJ, UK\\
\authormark{2}Instituto de Microelectrónica de Barcelona, IMB-CNM (CSIC), Campus UAB, 08193 Bellaterra, Barcelona, Spain}

\email{\authormark{*}F.Gardes@soton.ac.uk} 



\begin{abstract}
Photonic integrated circuits currently use platform intrinsic thermo-optic and electro-optic effects to implement dynamic functions such as switching, modulation and other processing. Currently, there is a drive to implement field programmable photonic circuits, a need which is only magnified by new neuromorphic and quantum computing applications. The most promising non-volatile photonic components employ phase change materials such as GST and GSST, which had their origin in electronic memory. However, in the optical domain, these compounds introduce significant losses potentially preventing a large number of applications. Here, we evaluate the use of two newly introduced low loss phase change materials, Sb\textsubscript{2}S\textsubscript{3} and Sb\textsubscript{2}Se\textsubscript{3}, on a silicon nitride photonic platform. We focus the study on Mach–Zehnder interferometers that operate at the O and C bands to demonstrate the performance of the system. Our measurements show an insertion loss below \SI{0.04}{\dB\per\um} for Sb\textsubscript{2}S\textsubscript{3} and lower than \SI{0.09}{\dB\per\um} for Sb\textsubscript{2}Se\textsubscript{3} cladded devices for both amorphous and crystalline phases. The effective refractive index contrast for Sb\textsubscript{2}S\textsubscript{3} on SiNx was measured to be 0.05 at 1310 nm and 0.02 at 1550 nm, whereas for Sb\textsubscript{2}Se\textsubscript{3},  it was 0.03 at 1310 nm and 0.05 at 1550 nm highlighting the performance of the integrated device.
\end{abstract}

\section{Introduction}
Photonic integrated circuits (PICs) have recently become an established and powerful technology that supports many applications \cite{chen2018emergence,smit2019past}. PICs are following the trend of integrating  electronic integrated circuits, using a range of components to manipulate photons and transfer information such as, on-chip optical waveguides \cite{khan2013silicon}, grating couplers \cite{hong2019high}, electro-optic modulators  \cite{brimont2011high,Thomson2012,faneca2019tuning}, photodetectors \cite{bie2017mote,xia2009ultrafast} and lasers on chip \cite{zhou2015chip}. Currently,  electronic circuits excel at digital computations, while photonics circuits are demonstrating tremendous progress at transporting and processing analogue and/or digital information at very high data rates \cite{miller2017silicon,chrostowski2019silicon,bogaerts2018silicon,bogaerts2020programmable}. Nowadays, PICs are commonly used in fibre-optic communications, but they are also useful in  applications in which light has a role, such as chemical, biological or spectroscopic sensors \cite{zinoviev2011integrated,wang2017iii}, metrology \cite{oton2016silicon}, and classical and quantum information processing \cite{perez2017multipurpose,miller2017silicon}. 

Programmable integrated photonics aim to provide a complementary approach to that based on application-specific photonic integrated circuits (ASPICs), which has been a growing research area over the last few years. The goal is to achieve in the optical domain similar advantages to those brought by field programmable gate arrays (FPGAs) over application specific integrated circuits (ASICs) in electronics \cite{perez2020principles}. Programmable integrated photonics has raised the attention of a range of emerging applications that not only demand flexibility and re-configurability but also  enable low-cost, compact and low-power consumption devices. To date, one of the biggest challenges faced by programmable silicon photonics is the integration of low-loss non-volatile components in applications such as quantum computing \cite{rudolph2017optimistic,qiang2018large,perez2017multipurpose}, microwave photonics \cite{bogaerts2020programmable}, neuromorphic computing \cite{nandakumar2018phase,boybat2018neuromorphic}, Internet of Things (IoT) \cite{szymanski2016securing}, and machine learning \cite{zhou2020passive}. 
Silicon nitride (SiNx) is one of the three currently commercially viable photonic platforms \cite{rahim2017expanding}, the other two being silicon \cite{absil2015silicon} and indium phosphide \cite{klamkin2018indium,zhao2019high}. SiNx provides an alternative low-cost CMOS compatible platform and similar to the silicon platform all fundamental non amplifying photonic components can be implemented \cite{Bucio2019}. The advantages over Si are fabrication flexibility, low temperature processing (<\SI{400}{\celsius}),  refractive index tunability, high transparency and low temperature sensitivity \cite{Bucio2018}. As a result, SiNx waveguides have been widely employed for light propagation in the mid infrared, the near infrared and in the visible range of the electromagnetic spectrum \cite{Milgram2007,Gaugiran2005a}. The versatility of the SiNx platform is key in the implementation of complex multi-layer photonic circuitry. The reduced mode confinement compared to Si, allows for compact active regions of programmable devices reducing the energy required to switch them \cite{baets2016silicon,li2020experimental}.
Phase change materials (PCMs) are being extensively studied for their use in photonic integrated circuits as they offer high refractive index contrast and  are non-volatile. Benefiting from prior art such as rewritable optical media and resistive memories, the mature technology of chalcogenide PCMs are now seen as a promising CMOS compatible route to provide the much needed non-volatile re-configurability in integrated photonics \cite{rios2018controlled}. PCMs have the ability to switch between two states, an amorphous and a crystalline phase with a resulting large optical contrast (allowing also intermediate states of crystallization). They are stable (years at room temperature) \cite{wuttig2017phase,wuttig2007phase,soref2018tutorial}, can be switched between states rapidly (nanoseconds or less) \cite{bruns2009nanosecond,ciocchini2016bipolar}, and exhibit high endurance (number of switching cycles) \cite{li2016reversible,cheng2011high}. The fundamental advantage of PCM-based programmable circuits compared to the conventional thermo-optic based approach is that energy is only consumed during the actual switching process. PCM-based devices have been demonstrated in a variety of applications such as switches \cite{stegmaier2017nonvolatile,de2019broadband}, wavelength division multiplexers \cite{feldmann2019integrated}, directional couplers \cite{xu2019low}, memories \cite{rios2015integrated,farmakidis2019plasmonic,feldmann2019integrated,gemo2019plasmonically} and neuromorphic devices \cite{cheng2017chip,feldmann2019all}.
Most PCM-based devices to date make use of the well-known pseudo-binary phase change material Ge\textsubscript{2}Sb\textsubscript{2}Te\textsubscript{5} (GST). Although it is stablished as a mature technology, GST was designed to offer fast switching (ns) and stability for applications that employ changes in reflectivity through the refractive index contrast, as applied in the CD, DVD, etc, or resistivity differences, applications which are largely independent on optical loss. On the other hand,  PICs that employ GST present strong coupling between amplitude and phase modulation due to the high absorption of the crystalline phase of the material. This coupling severely reduces the potential modulation schemes while it limits systems to a small number of components due to inherent losses.  

Emerging applications such as quantum information have been deemed viable ought to photonic integration but the vast computational requirements demand large systems prohibitive to lossy mediums. Even though GST-225 offers low losses \SI{0.039}{\dB\per\um} in the amorphous state,  the crystalline phase presents losses as high as \SI{2.7}{\dB\per\um} at \SI{1550}{\nm} \cite{faneca2020performance}. Addition of selenium to GST (GSST) presents an alternative solution to non-volatile low-loss phase change materials for photonic integrated circuits and different building blocks have been experimentally fabricated and tested. A low-loss directional coupler on a silicon platform has been demonstrated with losses as low as \SI{0.083}{\dB} \cite{de2019broadband}. Integrating GSST on a silicon nitride platform, a ring resonator switch has been experimentally measured in \cite{zhang2017broadband}  with losses of \SI{0.2}{\dB} and an extinction ratio of \SI{41}{\dB}. Also, a hybrid phase change material GSST-silicon Mach Zehnder modulator, with low insertion loss (\SI{3}{\dB}), to serve as node in a photonic neural network has been demonstrated in \cite{miscuglio2020artificial}. Finally, a family of broadband transparent optical phase change materials for high-performance non-volatile photonics was explored in \cite{zhang2019broadband}. These materials still maintain a relatively high loss (extinction coefficient) in order to switch the PCM optically. A novel family of low-loss PCMs which includes Sb\textsubscript{2}S\textsubscript{3} and Sb\textsubscript{2}Se\textsubscript{3} has been recently shown \cite{dong2019wide,delaney2020new}  and devices taking advantage of their optical properties such as Bragg gratings \cite{faneca2020chip} and ring resonators \cite{fang2020non} have been demonstrated. Sb\textsubscript{2}S\textsubscript{3} was first proposed in \cite{dong2019wide}, while \cite{delaney2020new}  introduced improvements by optimising the deposition conditions of Sb\textsubscript{2}S\textsubscript{3} and by achieving repeatable switching durability. More importantly, Delaney et al. \cite{delaney2020new} showed that Sb\textsubscript{2}Se\textsubscript{3}, in which sulfur is completely substituted by selenium, offers an equally low-loss alternative at lower switching temperatures. Sb\textsubscript{2}S\textsubscript{3} offers a refractive index contrast ($\Delta n$) between its states  of 0.60 at \SI{1550}{\nm} and 0.58 at \SI{1310}{\nm}, while for Sb\textsubscript{2}Se\textsubscript{3}, $\Delta n$=0.77 at \SI{1550}{\nm} and 0.82 at \SI{1550}{\nm}. Both materials present low inherent losses since their extinction coefficient, $k$ is less than \num{e-4} in both phases at \SI{1550}{\nm} and  at \SI{1310}{\nm}. The crystallisation temperature has been found to be \SI{290}{\celsius}  for Sb\textsubscript{2}S\textsubscript{3} and \SI{190}{\celsius} for Sb\textsubscript{2}Se\textsubscript{3}. Hence, Sb\textsubscript{2}Se\textsubscript{3} is easier to switch due to its lower crystallisation temperature while the   lower bandgap energy of \SI{1.48}{\eV} compared to the \SI{1.98}{\eV} of Sb\textsubscript{2}S\textsubscript{3} results in a shifted absorption spectrum towards higher wavelengths. As a result, the favourable properties of both materials cover a range of applications. In this paper, we integrate these novel low-loss PCM materials with a silicon nitride integrated platform to demonstrate their high performance in  phase modulation. We employ MZIs and use the effective refractive index contrast and absorption as metrics to assess their operation at two different wavelengths, \SI{1310}{\nm} and \SI{1550}{\nm}.

\section{Fabrication and characterization}
Silicon nitride Mach-Zehnder interferometers (MZIs) were fabricated exploiting the novel low loss phase change materials, Sb\textsubscript{2}S\textsubscript{3} and Sb\textsubscript{2}Se\textsubscript{3}  to utilize them as  non-volatile phase modulators building blocks. The MZIs building block structures were designed in order to test the performance of both Sb\textsubscript{2}S\textsubscript{3} and Sb\textsubscript{2}Se\textsubscript{3} incorporated in the proposed silicon nitride platform at both \SI{1310}{\nm}  and \SI{1550}{\nm} wavelengths.  These devices were fabricated on 8" Si wafers with a \SI{2}{\um} thermally-grown SiO\textsubscript{2} on top of which a \SI{300}{\nm} SiNx layer (n = 2.0) was deposited at \SI{350}{\celsius} using the NH\textsubscript{3} PECVD process fully detailed in \cite{DominguezBucio2017}. The test structures were defined on the wafers by means of \SI{248}{\nm} deep-UV (DUV) lithography and the pattern was transferred onto the SiN layer using inductively coupled plasma etching (ICP) with an etch depth of \SI{300}{\nm} forming ridge structures.  The PCMs were deposited through windows defined by the second DUV lithography step. The PCM deposition was performed by RF sputtering as described in \cite{delaney2020new}. A \SI{10}{\nm} ZnS/SiO\textsubscript{2} capping layer was deposited by sputtering right after the deposition of the PCM. The photoresist was lift-off by sequentially dipping in room temperature acetone and   \SI{80}{\celsius} NMP with ultrasonic agitation. After rinsing  in acetone, IPA and drying with a nitrogen gun a second \SI{10}{\nm} layer of ZnS/SiO\textsubscript{2} was deposited to provide further protection of the exposed interfaces.

\begin{figure}[htbp]
\centering
\includegraphics[width=0.55\textwidth]{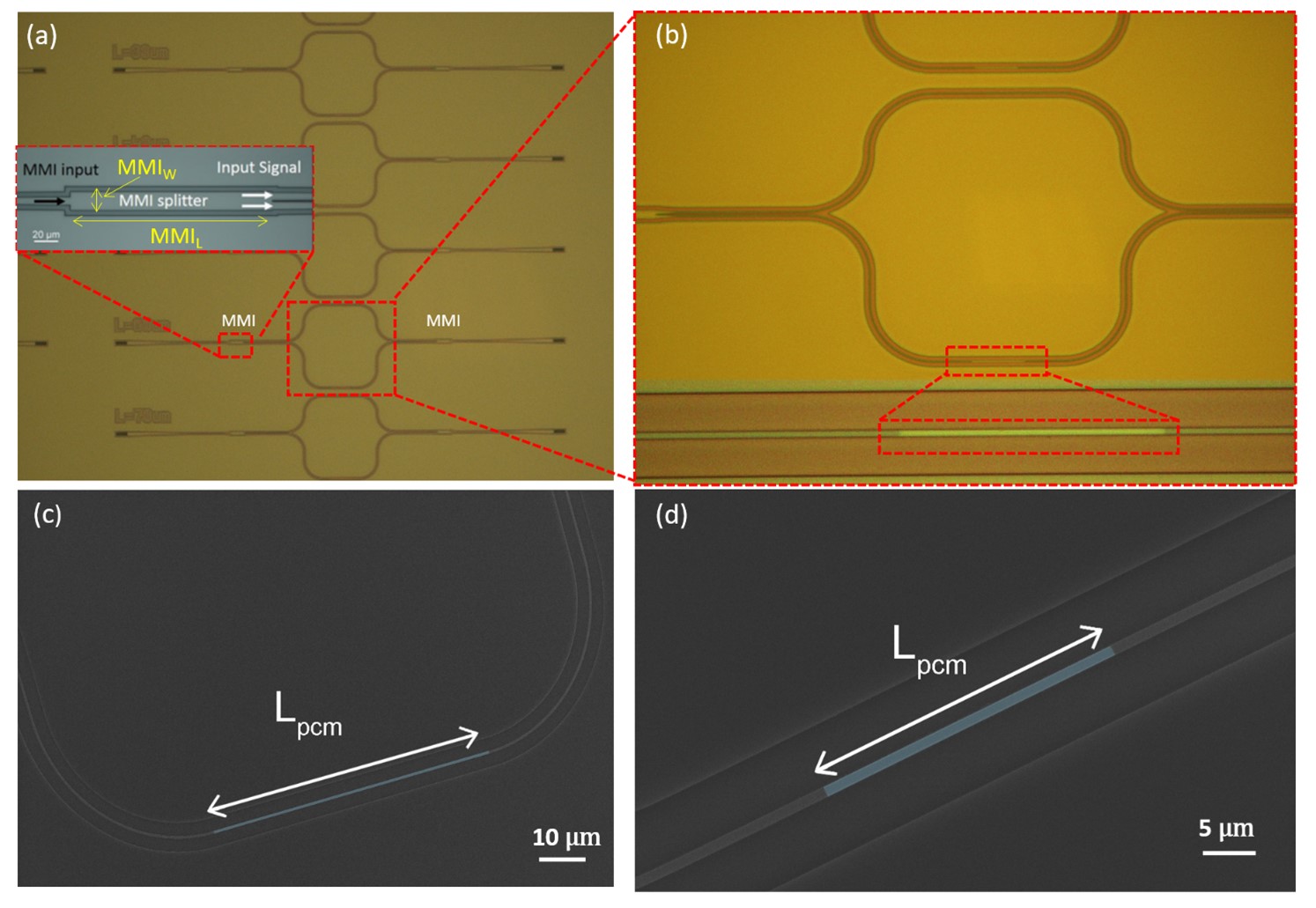}
\caption{ Optical microscope and SEM images of the fabricated structures. (a) MZIs using different cell lengths. (b) Zoom in image of (a)  with an inset of the PCM deposited layer. (c) SEM image of the longer arm of the MZI with the cell deposited on top. (d) Zoom in SEM image of a PCM cell with fake color\label{SEM}}
\end{figure}

The  layout we designed included a set of 12 MZIs with single-mode waveguide cross-sections of  \SI{1.2x0.3}{\square\um} for  \SI{1550}{\nm} (C-band) and \SI{0.9x0.3}{\square\um} for  \SI{1310}{\nm} (O-band). These MZIs have an optical path length difference of $\Delta L$= \SI{60}{\um} at \SI{1310}{\nm}  and $\Delta L$= \SI{40}{\um} at \SI{1550}{\nm}  between their arms, see Fig. \ref{SEM}(a). In the longer arm of the MZI, PCM cells with a thickness of \SI{20\pm 4}{\nm} for Sb\textsubscript{2}S\textsubscript{3} and  \SI{15\pm 4}{\nm} for Sb\textsubscript{2}Se\textsubscript{3}  and different lengths ranging from \SIrange{2}{125}{\um} were deposited. To split and combine the light, \SI{3}{\dB} broadband multi-mode interferometers (MMIs) were employed, see inset of Fig \ref{SEM}(a). The dimensions of the MMI at $\lambda $ = \SI{1.31}{\um}, were MMI\textsubscript{L} = \SI{41}{\um} and MMI\textsubscript{W} = \SI{9}{\um}, and respectively for $\lambda $ = \SI{1.55}{\um}, MMI\textsubscript{L} = \SI{64}{\um} and MMI\textsubscript{W} = \SI{11}{\um}. SEM images of the longer MZI arm were taken with different lengths of the PCM cell. In Figure \ref{SEM}(c), a cell length of a \SI{100}{\um} was captured and in Fig. \ref{SEM}(d) a zoom in image of the longer arm with a cell length of \SI{20}{\um} was measured. 
All the devices included input and output grating couplers (GCs) designed to couple the light at either of both wavelengths of interest for characterisation, \SI{1310} or \SI{1550}{\nm}. The GCs designed for \SI{1310}{\nm} consisted of a \SI{10}{\um} x \SI{37}{\um} surface grating with a period of \SI{1038}{\nm} tapered down to a single-mode waveguide width of \SI{900}{\nm}, whereas the ones designed for \SI{1550}{\nm} had a surface grating with the same dimensions and a period of \SI{1238}{\nm} tapered to a single-mode width of \SI{1200}{\nm}. The angle between the optical fibers delivering light and the gratings was selected to be \ang{10} or \ang{15} to the normal to ensure maximum coupling at \SI{1310} or \SI{1550}{\nm}, respectively. Additionally, the layout included  separate structures with two GCs connected back-to-back for normalisation purposes.
The spectral response for all the devices was characterised using two different tunable laser sources. An Agilent 8163B Lightwave Multimeter for the response at \SI{1550}{\nm} and a similar Agilent 8164B Lightwave Measurement System for the response at \SI{1310}{\nm}. The measurements were first performed for the amorphous state of the PCMs and then for their crystalline state. The phase change was induced thermally by heating the chip on a hot plate at the crystallisation temperature for 10 minutes. In both cases, the polarization of the light was controlled to ensure that only TE modes could propagate through the devices and the measurements were normalized to extract the loss contribution of the grating couplers.

\begin{figure*}[hbtp]
\centering
\includegraphics[width=0.85\textwidth]{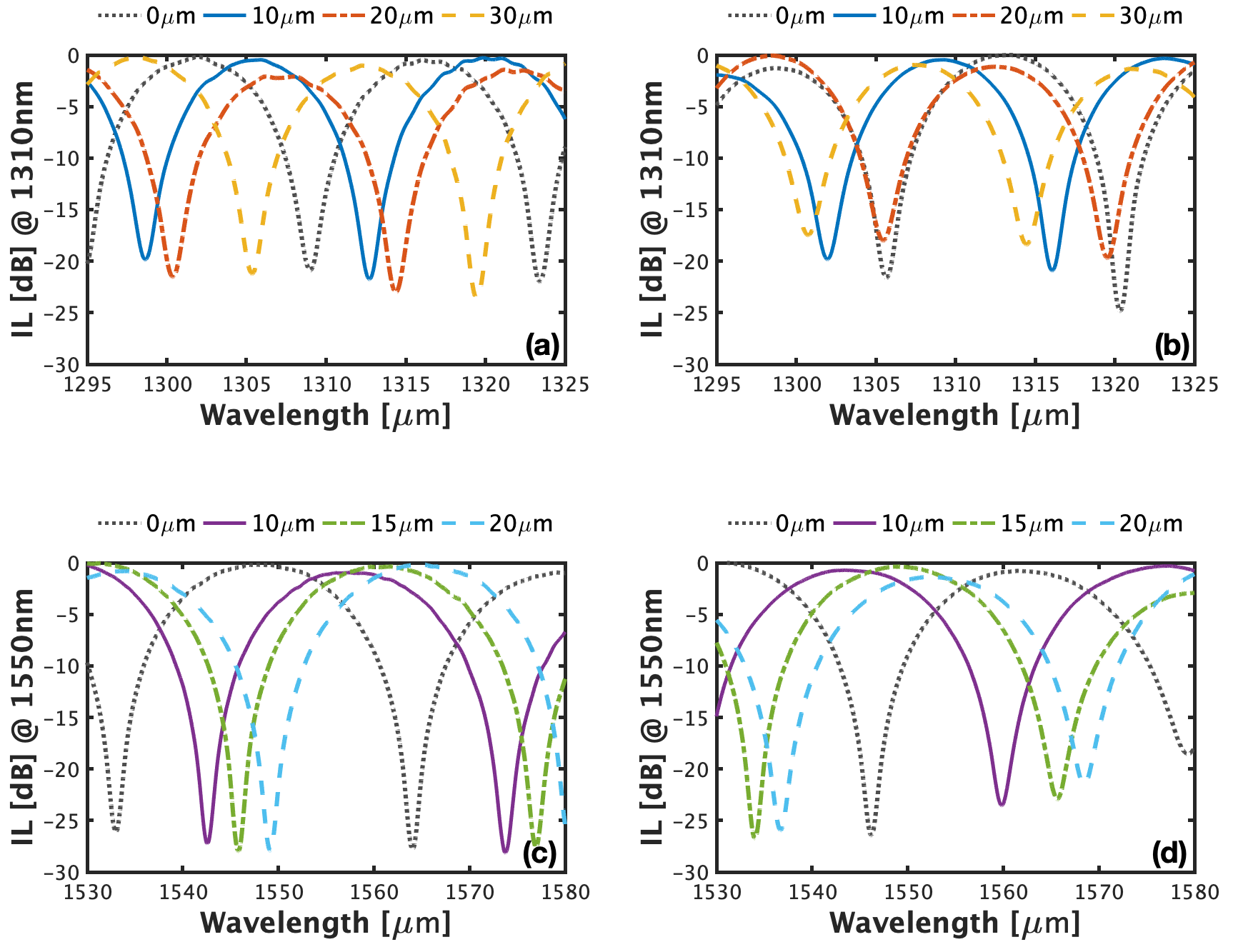}
\caption{MZI response for different cell lengths, as indicated in the legends, of Sb\textsubscript{2}S\textsubscript{3} in the (a) amorphous state at \SI{1310}{\nm}, (b) crystalline state at \SI{1310}{\nm}, (c) amorphous state at \SI{1550}{\nm} and (d) crystalline state at \SI{1550}{\nm}.\label{MZI_Sb2S3}}
\end{figure*}

\section{Results and discussion}

Silicon nitride MZIs were fabricated using  novel low loss phase change materials, Sb\textsubscript{2}S\textsubscript{3} and Sb\textsubscript{2}Se\textsubscript{3} to demonstrate a low-loss non-volatile MZI building block for future electro-refractive non-volatile photonic integrated circuit applications in the O and C-bands. Firstly, the Sb\textsubscript{2}S\textsubscript{3} material is analysed. The bare MZI structure without phase change material was characterized and normalized with respect to GCs connected back-to-back. Afterwards, the response of the MZIs for different lengths of the PCM cell were characterized in both amorphous (Fig. \ref{MZI_Sb2S3}(a) and (c)) and crystalline states (Fig. \ref{MZI_Sb2S3}(b) and (d)) at both target wavelengths.

In Fig. \ref{MZI_Sb2S3}(a), the bare MZI presents a free-spectral range (FSR) of \SI{13}{\nm} in the O-band. Only cell lengths of 10, 20 and \SI{30}{\um} were selected to be plotted in order to  not overload the graph with all the cell lengths used in the study. The shift in wavelength ($\Delta\lambda$) produced by the cell lengths were 2.75, 4.38 and \SI{9.47}{\nm} respectively in the amorphous state respect to the bare MZI. In Fig. \ref{MZI_Sb2S3}(b), the crystalline state is shown in the O-band. For the bare structure after crystallization, the FSR maintained the previous value (prior to crystallization), even though a shift has been produced  in the dips of the optical spectrum due to the crystallization process to which the chip was submitted. In this case, a shift in wavelength for the cells of length 10, 20 and \SI{30}{\um}  of 7.64, 9.5 and \SI{13.03}{\nm}, respectively,  was measured.

In each case, an extinction ratio (ER) higher than \SI{20}{\dB} is shown for the different MZI structures. 
Fig \ref{MZI_Sb2S3}(c) represents the MZI response using Sb\textsubscript{2}S\textsubscript{3} in the amorphous state of the PCM in the C-band. A FSR of \SI{31}{\nm} was extracted, higher than the one measured in  the O-band (\SI{13}{\nm}), a shift in wavelength for cell lengths of 10, 15 and \SI{20}{\um} of 9.67, 12.76 and \SI{16.07}{\nm} was experimentally demonstrated.  In Fig \ref{MZI_Sb2S3}(d), the crystalline state of the PCM for the C-band is shown and shifts of 13.5, 19.5 and \SI{22.5}{\nm} were obtained for the three different cell lengths, 10, 15 and \SI{20}{\um}. ERs as high as \SI{30}{\dB} were experimentally demonstrated in this range of the spectrum, achieving a difference of 7.64, 9.5 and \SI{13.03}{\nm} for the O-band and respectively 3.83, 6.74 and \SI{6.43}{\nm} for the C-band between crystalline and amorphous dips. 

\begin{figure}[bp]
\centering
\includegraphics[width=0.85\textwidth]{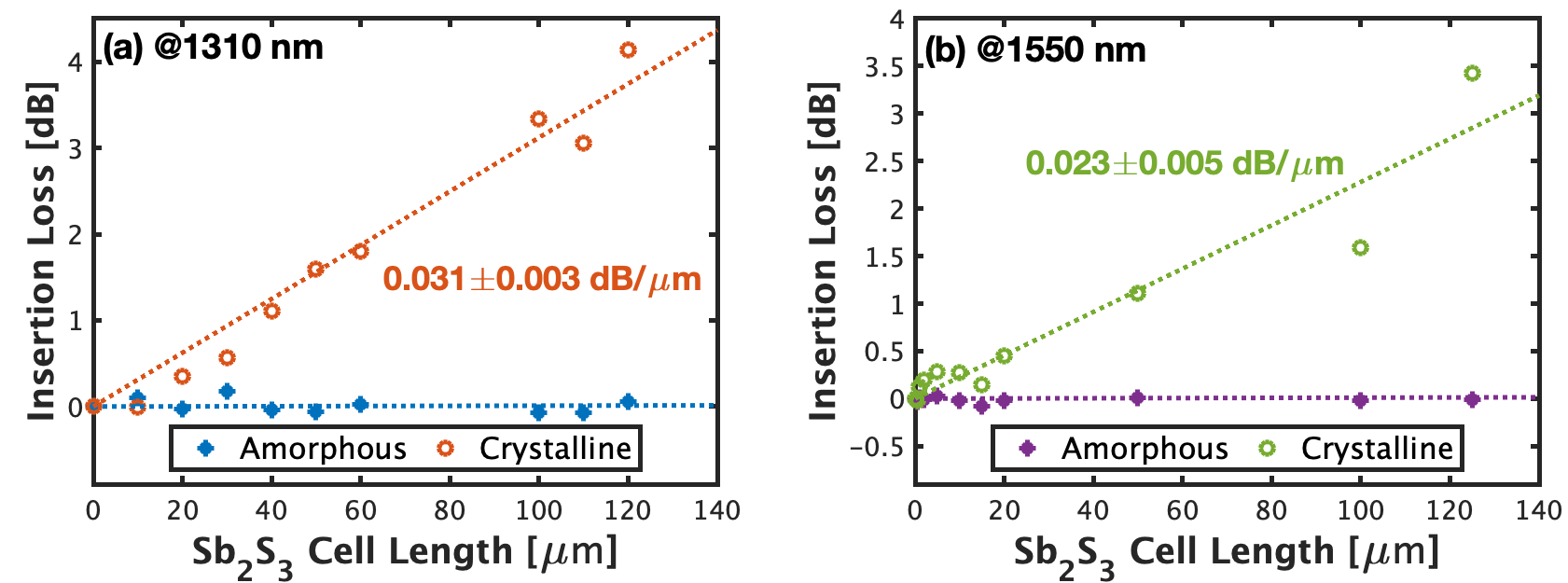}
\caption{ Measured losses  for different cell lengths of Sb\textsubscript{2}S\textsubscript{3} in both amorphous and crystalline state at  (a)  \SI{1310}{\nm} and  (b) \SI{1550}{\nm}.\label{PL_Sb2S3}}
\end{figure}

Afterwards, we characterized  the insertion loss (IL) introduced by the PCM layer ($\alpha$) and the modulation in effective refractive index ($\Delta n_{eff}$)   in its amorphous and crystalline states. The IL and the modulation in phase were extracted from the spectral response of the fabricated MZIs. To characterize the losses of the PCM cell, the ER of each individual MZI with the corresponding cell length was evaluated and consequently the loss coefficient extracted using the equation from Ref.\cite{rios2014chip}:

\begin{equation}
    ER=\left(\frac{1+e^{\frac{-\alpha L_{pcm}}{2}}}{1-e^{\frac{-\alpha L_{pcm}}{2}}}\right)^2
\end{equation}

\noindent where $L_{pcm}$ is the length of the PCM cell.  For the amorphous state of the PCM, the ER was not varying significantly while increasing the cell length, showing a low loss introduced by the PCM in both ranges of the spectrum. The extracted losses in the amorphous state for the wavelengths of interest are lower than \SI{e-4}{\dB\per\um}. For the crystalline state, ERs were decreasing with increasing the cell lengths, showing an increment in losses in the crystalline state respect to the amorphous state. The measured IL   in the crystalline state, were as low as \SI{0.031\pm0.003}{\dB\per\um} and \SI{0.023\pm0.005}{\dB\per\um} at 1310 nm and \SI{1550}{\nm} respectively, see Fig. \ref{PL_Sb2S3}. 

\begin{figure}[bp]
\centering
\includegraphics[width=0.85\textwidth]{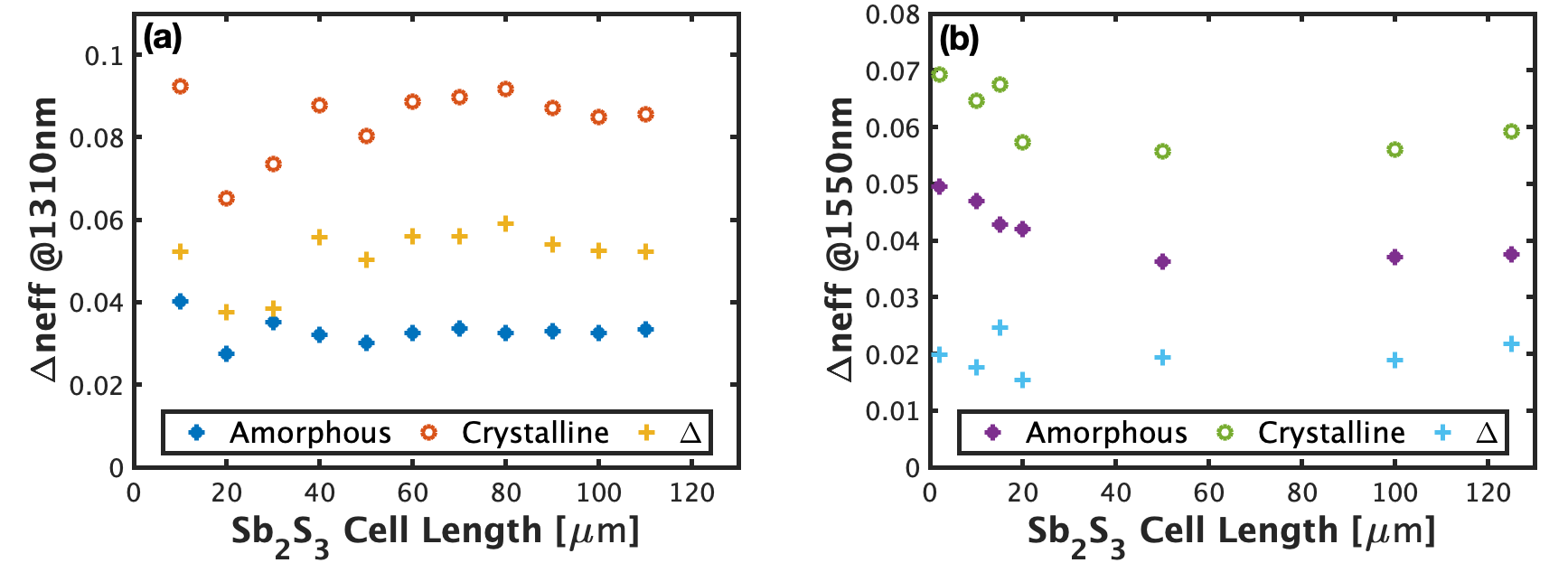}
\caption{ Effective refractive index difference ($\Delta_{n_{eff}}$) as a function of cell length for both amorphous and crystalline states of Sb\textsubscript{2}S\textsubscript{3} at (a) \SI{1310}{\nm} and (b) \SI{1550}{\nm}.\label{Dneff_Sb2S3}}
\end{figure}

Once the IL of the PCM was characterized, the effective refractive index modulation of the material can be obtained using the equation from Ref.\cite{zhang2017ultracompact}:

\begin{equation}
    \Delta n_{eff}^{(a-c)}= \frac{\lambda}{FSR}\cdot \frac{\Delta \lambda^{(a-c)}}{L_{pcm}}
\end{equation}

\noindent where  $\Delta n_{eff}$ is the variation produced in the effective refractive index when the PCM cell is introduced compared to the bare structure (no PCM), the superscripts $a$ and $c$ refer to the amorphous and crystalline states respectively, FSR is the free spectral range as previously introduced, and $\Delta\lambda$  is the shift in the dip of the amorphous and crystalline states with respect to the dip produced by the bare MZI waveguide as introduced previously. 
Figure \ref{Dneff_Sb2S3} shows the $\Delta n_{eff}$ for both ranges of the spectrum,  \SI{1310}{\nm} (Fig. \ref{Dneff_Sb2S3}(a)) and  \SI{1550}{\nm} (Fig. \ref{Dneff_Sb2S3}(b)). The modulation between the amorphous state and the bare waveguide, the crystalline state and the bare waveguide and the amorphous state compared with the crystalline state ($\Delta$) are presented in Fig. \ref{Dneff_Sb2S3}. The overall effective refractive index contrast between amorphous and crystalline states was measured to be 0.05 and 0.02 at  \SI{1310}{\nm} and  \SI{1550}{\nm}, respectively.

\begin{figure*}[htbp]
\centering
\includegraphics[width=0.85\textwidth]{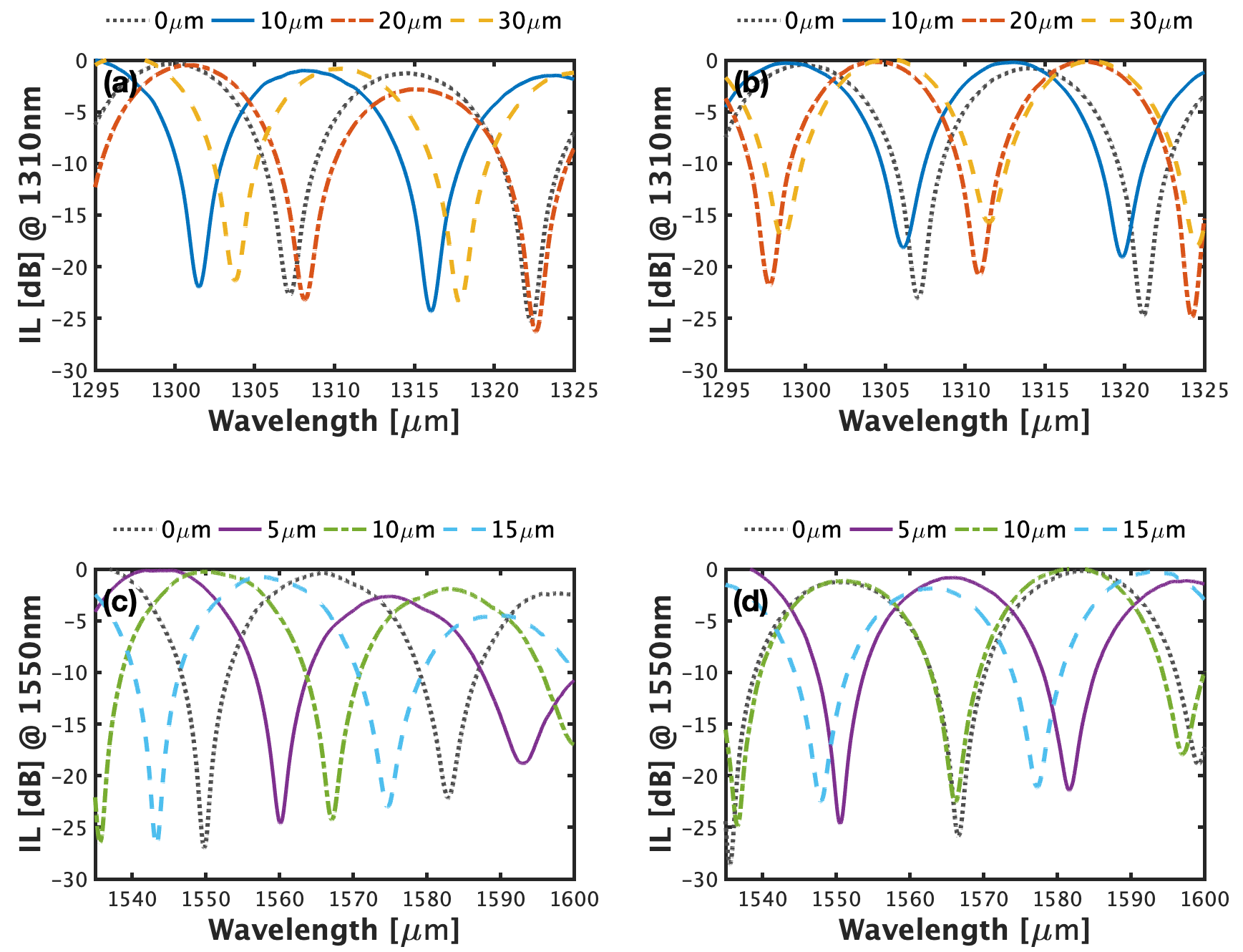}
\caption{MZI response for different cell lengths, as indicated in the legends, of Sb\textsubscript{2}Se\textsubscript{3} in the (a) amorphous state at \SI{1310}{\nm}, (b) crystalline state at \SI{1310}{\nm}, (c) amorphous state at \SI{1550}{\nm} and (d) crystalline state at \SI{1550}{\nm}.\label{MZI_Sb2Se3}}
\end{figure*}

The same procedure was used in order to characterize the building block performance of the MZIs devices for the material, Sb\textsubscript{2}Se\textsubscript{3}. In this case, the spectral response for the different MZIs is presented in Fig. \ref{MZI_Sb2Se3}. Fig \ref{MZI_Sb2Se3}(a) shows the amorphous state of the Sb\textsubscript{2}Se\textsubscript{3} material in the O-band. A shift of 9, 15.5 and \SI{25.4}{\nm} between the bare  MZI dip and the MZI dip with  PCM cells of 10, 20 and \SI{30}{\um}  was measured respectively. For the crystalline state, an increase in the shift is produced compared with the amorphous state and  a shift of 13, 17.9 and  \SI{31.5}{\nm} is measured for cell lengths of  10, 20 and \SI{30}{\um} respectively. In all the cases ERs higher than 20 dB were demonstrated, see Fig. \ref{MZI_Sb2Se3}(b). 
In the C-band, the Sb\textsubscript{2}Se\textsubscript{3}  material showed a modulation in wavelength of 10.3, 17.5 and \SI{24.9}{\nm} for PCM amorphous cell lengths of 5, 10 and \SI{15}{\um}. When switching to crystalline state, the modulation is 14.9, 30.65 and \SI{41.4}{\nm} respectively, see Fig. \ref{MZI_Sb2Se3}(c-d). Resulting in a $\Delta \lambda  $ of 4, 2.1 and \SI{6.1}{\nm} for O-band and respectively 4.6, 13.15 and \SI{16.5}{\nm} for the C-band.

\begin{figure}[htbp]
\centering
\includegraphics[width=0.85\textwidth]{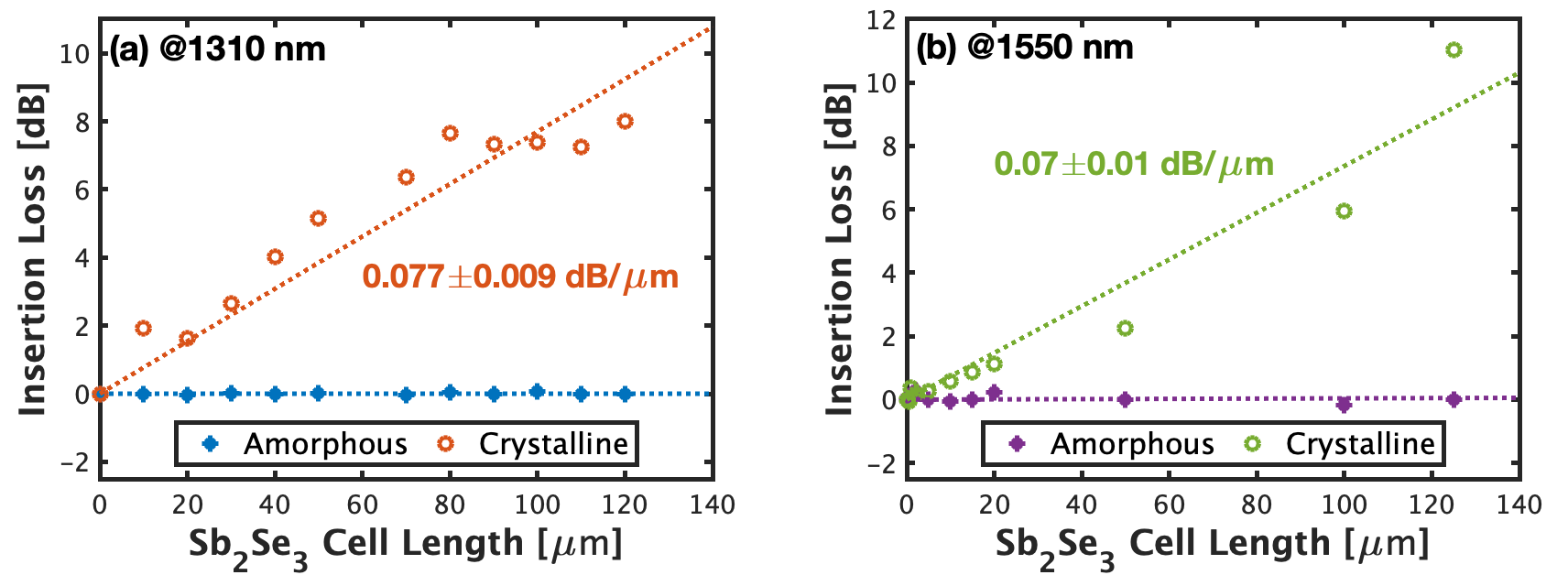}
\caption{ Measured losses  for different cell lengths of Sb\textsubscript{2}Se\textsubscript{3} in both amorphous and crystalline state at  (a)  \SI{1310}{\nm} and  (b) \SI{1550}{\nm}.\label{PL_Sb2Se3}}
\end{figure}

\begin{figure}[htbp]
\centering
\includegraphics[width=0.85\textwidth]{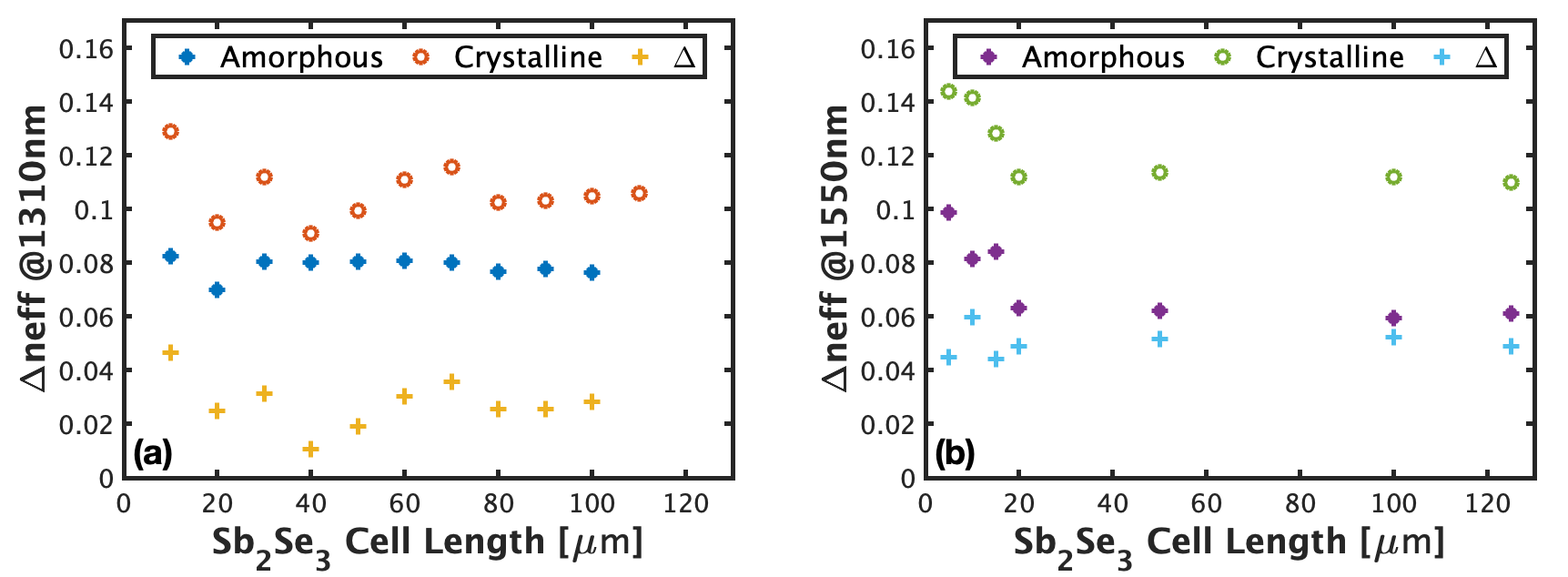}
\caption{ Effective refractive index difference ($\Delta_{n_{eff}}$) as a function of cell length for both amorphous and crystalline states of Sb\textsubscript{2}Se\textsubscript{3} at (a) \SI{1310}{\nm} and (b) \SI{1550}{\nm}.\label{Dneff_Sb2Se3}}
\end{figure}

The ILs and the effective index modulation were extracted using the same method used for the Sb\textsubscript{2}S\textsubscript{3} material. In this case, the measured losses in both states are shown in Fig. \ref{PL_Sb2Se3}. In the amorphous state of  Sb\textsubscript{2}Se\textsubscript{3} for the wavelengths of interest, losses are lower than \SI{e-4}{\dB\per\um}, whereas in the crystalline state, losses are \SI{0.077\pm0.009}{\dB\per\um} at \SI{1310}{\nm}  and \SI{0.07\pm0.01}{\dB\per\um} at \SI{1550}{\nm}. 
The effective refractive index difference between the amorphous and crystalline state was measured to be 0.03 at \SI{1310}{\nm} and 0.05 at \SI{1550}{\nm} as shown in Fig. \ref{Dneff_Sb2Se3}.

Table \ref{table2} presents the optical performance of the currently popular  phase change materials GST, GSST and the novel materials discussed in this paper, Sb\textsubscript{2}S\textsubscript{3} and Sb\textsubscript{2}Se\textsubscript{3} on our integrated silicon nitride platform. In the table, a comparison of the cell length required in order to achieve a $\pi$ shift ($L_{\pi}=\lambda/2\cdot(\Delta n_{eff})$) and the IL that this cell introduces in the device are also shown.
\noindent GST offers a $\pi$ phase shift at the shortest length while GSST requires double, Sb\textsubscript{2}Se\textsubscript{3} triple and Sb\textsubscript{2}S\textsubscript{3} seven times that length (the GST length) for the same shift at \SI{1550}{\nm}. However, the popularity of GSST stems from the three times lower insertion loss when compared to GST. Our Sb\textsubscript{2}Se\textsubscript{3} and Sb\textsubscript{2}S\textsubscript{3} films offer a stark improvement with the insertion loss at 13 and 18 times lower than GST respectively, significantly surpassing the benefits of GSST.  In the O-band (\SI{1310}{\nm}), the required length for a $\pi$ shift ($L_{\pi}$)  in the case of   Sb\textsubscript{2}Se\textsubscript{3} is \SI{21.83}{\um}, although, there is a reduction of \SI{12.5}{\um}  for GST, \SI{14.5}{\um}  for GSST and \SI{9}{\um} for the Sb\textsubscript{2}S\textsubscript{3} material. Again in this range of the spectrum the IL introduced by the novel materials highly improve the performance of GST and GSST. It is therefore obvious that both materials provide in the O band and C band an incontestable improvement on insertion losses compared with GST and GSST in the crystalline state. 

\begin{table}[htbp]
\begin{center}
\begin{adjustbox}{max width=\textwidth}
\begin{tabular}{|c||c|c|c|c|c||c|c|c|c|c|}
\hline
&\multicolumn{5}{|c||}{C-band @ \SI{1550}{\nm}}&\multicolumn{5}{|c|}{O-band @ \SI{1310}{\nm}} \\\hline\hline

\multirow{2}{*}{PCM}& $\alpha_c$ &$\alpha_a$ & \multirow{2}{*}{$\Delta n_{eff}$}&$L_{\pi}$  &IL$_{\pi}$ &$\alpha_c$ &$\alpha_a$ & \multirow{2}{*}{$\Delta n_{eff}$}&$L_{\pi}$  &IL$_{\pi}$\\
&[\si{\dB\per\um}]&[\si{\dB\per\um}]&&[\si{\um}]&[\si{\dB}]&[\si{\dB\per\um}]&[\si{\dB\per\um}]&&[\si{\um}]&[\si{\dB}]\\
\hline
GST  \cite{faneca2020performance}&2.860&0.039&0.14&5.53&15.81&6.660&0.228&0.07&9.36& 62.4\\
\hline
GSST* \cite{faneca2020chip} &0.53&<\num{e-4}&0.074&10.5&5.56&1.42&<\num{e-4}&0.09&7.28&10.3\\
\hline
Sb\textsubscript{2}S\textsubscript{3} &\num{0.023+-0.005}&<\num{e-4}&0.02&38.75&0.89&\num{0.031+-0.003}&<\num{e-4} &0.05&13.1&0.39\\
\hline

Sb\textsubscript{2}Se\textsubscript{3}&\num{0.07+-0.01}&<\num{e-4}&0.05&15.5&1.15&\num{0.077+-0.009}&<\num{e-4} &0.03&21.83&1.68\\

\hline

\end{tabular}
\end{adjustbox}
\caption{The device performance comparison between the  C-band (\SI{1550}{\nm}) and the O-band (\SI{1310}{\nm})  for  different phase change materials: GST \cite{faneca2020performance}, GSST \cite{faneca2020chip}, Sb\textsubscript{2}S\textsubscript{3} and Sb\textsubscript{2}Se\textsubscript{3}.*Refers to a simulated value. }
\label{table2}
\end{center}
\end{table}

\section{CONCLUSIONS}
In this paper, a MZI building block is demonstrated exploiting a novel family of low loss phase change materials  in the telecommunications O and C-bands for future non-volatile electro-refractive modulation in programmable photonic integrated circuits in mid index waveguides. Losses lower than  \SI{0.09}{\dB\per\um} for Sb\textsubscript{2}Se\textsubscript{3} and  \SI{0.04}{\dB\per\um} for Sb\textsubscript{2}S\textsubscript{3} are demonstrated in both ranges of the spectrum with changes in effective refractive index higher than 0.02.
This study thus provides useful characterization information for the operation of non-volatile integrated photonic circuits based on phase change materials in two of the most important spectral ranges used for optical communications. The demonstration of these low loss non volatile building blocks, using a low loss and back end of the line compatible waveguide platform, have opened up  pathways for more complex PIC's architecture and extended applications.

\section*{Funding}
Horizon 2020 Framework Programme (871391); Engineering and Physical Sciences Research Council (EP/M015130/1, EP/T007303/1, EP/R003076/1, EP/N00762X/1).

\section*{Disclosures}
The authors declare no conflicts of interest.

\bibliography{sample}






\end{document}